\documentclass[conference]{IEEEtran}
\IEEEoverridecommandlockouts
\usepackage{cite}
\usepackage{amsmath,amssymb,amsfonts}
\usepackage{algorithmic}
\usepackage{graphicx}
\usepackage{textcomp}
\usepackage{xcolor}

\usepackage{color}

\begin{document}

\title{Electrocardiogram Generation and Feature Extraction Using a Variational Autoencoder
\thanks{The work is supported by the Ministry of Education and Science of Russian Federation 
(project 14.Y26.31.0022).}}

\author{\IEEEauthorblockN{V. V. Kuznetsov \qquad V. A. Moskalenko \qquad N. Yu. Zolotykh}
\IEEEauthorblockA{Institute of Information Technologies, Mathematics and Mechanics\\
Lobachevsky State University of Nizhni Novgorod\\
Nizhni Novgorod, Russia\\
Email: \{vladislav.kuznetsov,viktor.moskalenko,nikolai.zolotykh\}@itmm.unn.ru}}


\maketitle

\begin{abstract}
We propose a method for generating an
electrocardiogram (ECG) signal for one cardiac cycle using a variational autoencoder. Using this method we extracted a vector of new $25$ features, which in many cases can be interpreted. The generated ECG has quite natural appearance. The low value of the Maximum Mean Discrepancy metric, $3.83\times 10^{-3}$, indicates good quality of ECG generation too. The extracted new features will help 
to improve the quality of automatic diagnostics of cardiovascular diseases. Also, generating new synthetic ECGs will allow us to solve the issue of the lack of labeled ECG for use them in supervised learning.
\end{abstract}


\section{Introduction}

All the experience gained by the machine learning community shows that the quality of the decision rule largely depends on what features of samples are used.
The better the feature description, the more accurately the problem can be solved. Typically, the features require their interpretability, since it means the adequacy of the features to the real-world problem.

The traditional way to build a good feature description was to use an expert knowledge. Specialists in a particular subject area offer various methods for constructing the feature descriptions, which are then tested in solving practical problems. Another approach for constructing a good feature description is automatic feature extraction (also called dimensionality reduction).

There is a lot of methods for automatic feature extraction, such as principal component analysis, independent component analysis, principal graphs and manifolds, kernel methods, autoencoders, embeddings etc.
Among the most powerful and perspective approaches, we mention principal graphs and manifolds \cite{GorbanKeglWunschZinovyev2008} and methods used deep learning \cite{LeCunBengioHinton2015,GoodfellowBengioCourville2016}.

Here we examine a method for automatic feature extraction, so called variational autoencoder (VAE) \cite{KingmaWelling2013,Doersch2016}, for the problem of automatic electrocardiogram (ECG) processing.  

\begin{figure}[b]
    \centering
	\includegraphics*[width=.75\linewidth]{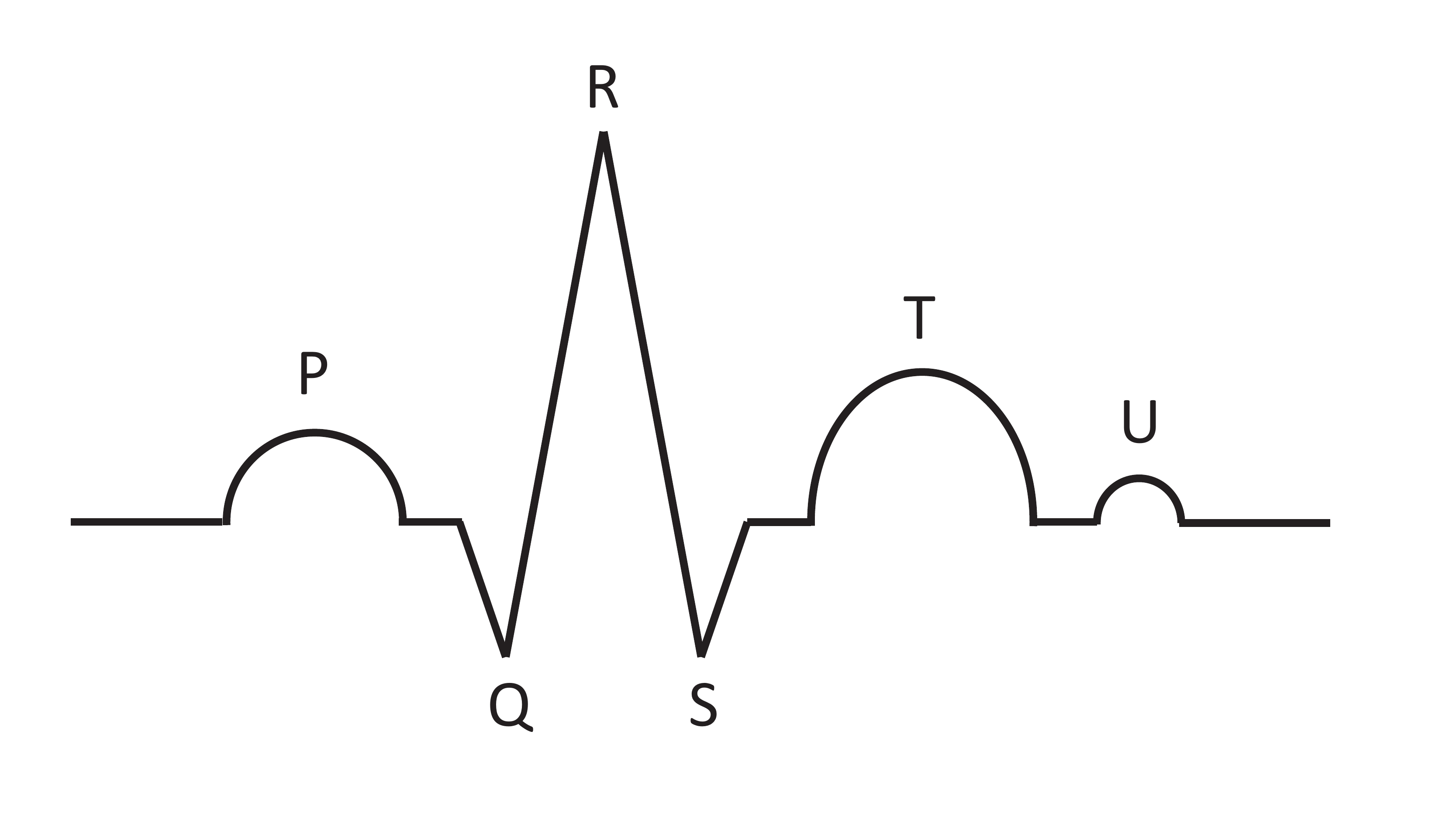}\\
	\caption{Schematic representation of main parts of the ECG signal for one cardiac cycle: P, T, U waves and QRS complex, consisting of Q, R and S peaks.}
	\label{fig_cardio_cycle}
\end{figure}

The electrocardiogram is a record of the electrical activity of the heart, obtained with the help of electrodes located on the human body. Electrocardiography
is one of the most important methods in cardiology. Schematic representation of the main part of ECG is shown in Figure \ref{fig_cardio_cycle}.  One cardiac cycle (the performance of the heart from the beginning of one heartbeat to the beginning of the next) contains P, T, U waves and QRS complex, consisting of Q, R and S peaks. The size, shape, location of these parts gives great diagnostic information about the work of the heart and about the presence/absence of certain diseases.

Recently, machine learning (especially deep learning) methods are widely used for automatic ECG analysis.
See the recent review \cite{HongZhouShangXiaoSun2019}.
The application tasks include ECG segmentation, disease detection, sleep staging, biometric human identification, denoising, and the others \cite{HongZhouShangXiaoSun2019}. A variety of classical and new methods are used. Among them are discriminant analysis, decision trees, support vector machine, fully-connected and convolutional neural networks, recurrent neural networks, generative adversarial networks, autoencoders etc \cite{SchlapferWellens2017,HongZhouShangXiaoSun2019}.

From our point of view, the most interesting and fruitful directions in applying deep learning methods to ECG analysis is the generation of synthetic ECG and automatic extraction of new interpretable features. The problem of ECG generation is devoted to several works \cite{ZhuYeFuLiuShen2019,DelaneyBrophyWard2019,GolanyRadinsky2019}. The authors of those papers used different variations of generative adversarial networks (GANs)
\cite{Goodfellowet2014}.  The best results concerning the ECG generation were obtained in \cite{DelaneyBrophyWard2019}. The authors report on the Maximum Mean Discrepancy (MMD) metric equals to $1.05\times 10^{-3}$. 

Our approach to generate ECG is based on VAE. We propose a neural network architectures for an encoder and a decoder for generating synthetic ECGs and extracting new features. The generated synthetic ECGs look quite natural. MMD equals to $3.83\times 10^{-3}$, which is worse than the value obtained in \cite{DelaneyBrophyWard2019}, but we note that the comparison of these two metric values is not very correct, since the values were obtained on different training tests and for solving similar, but different problems.

The main advantage of our work is that we propose the method for extracting new features. Our experiments show that these features are quite interpretable. This fact allows us to hope that using these features will help to improve the quality of automatic diagnostics of cardiovascular diseases. Also, generating new synthetic ECGs will allow us to fix the issue of the lack of labeled ECG for use them in supervised learning.

\begin{figure*}
\centering
\includegraphics[width=1\textwidth]{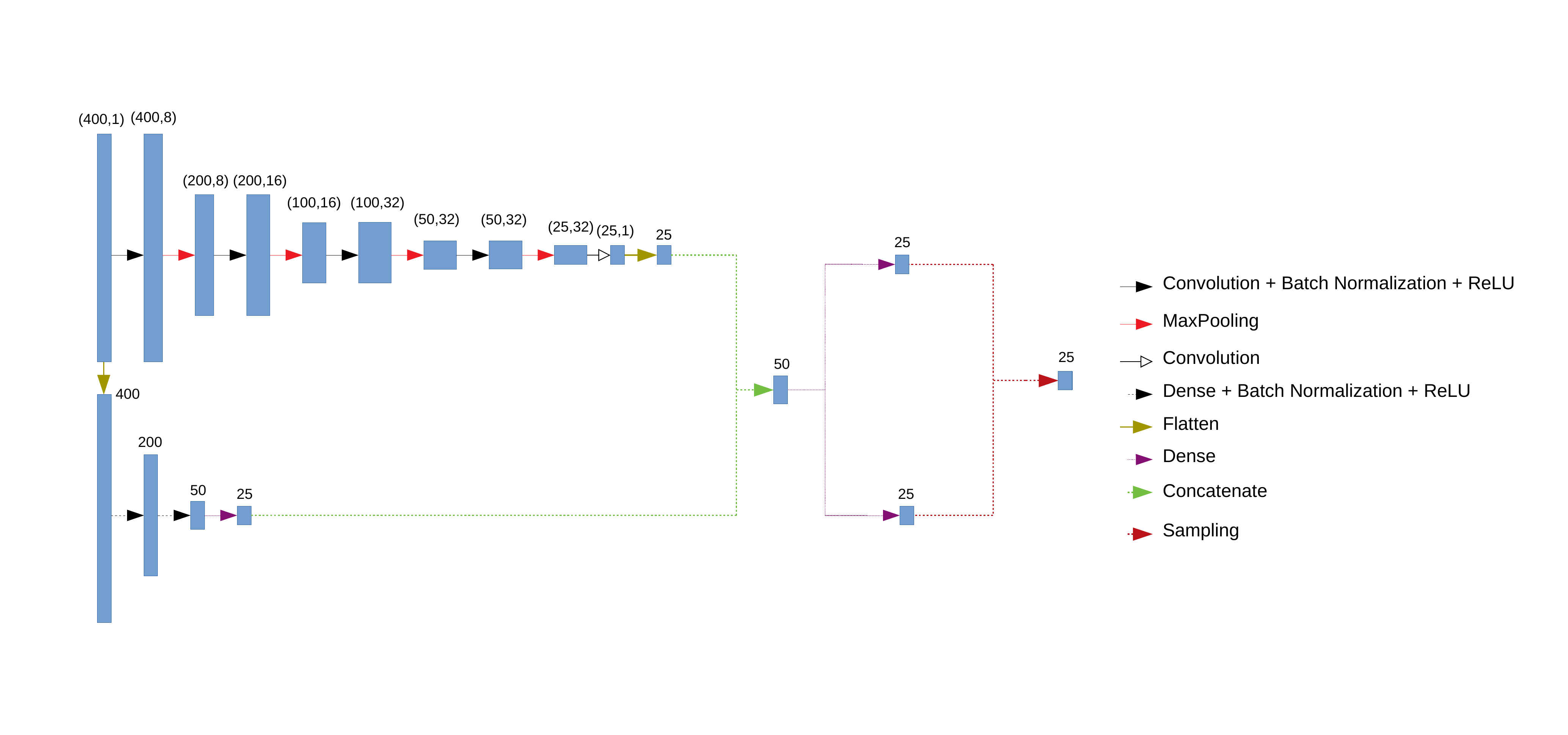}
\caption{Encoder architecture}\label{pic_encoder}
\end{figure*}

\begin{figure*}
\centering
\includegraphics[width=1\textwidth]{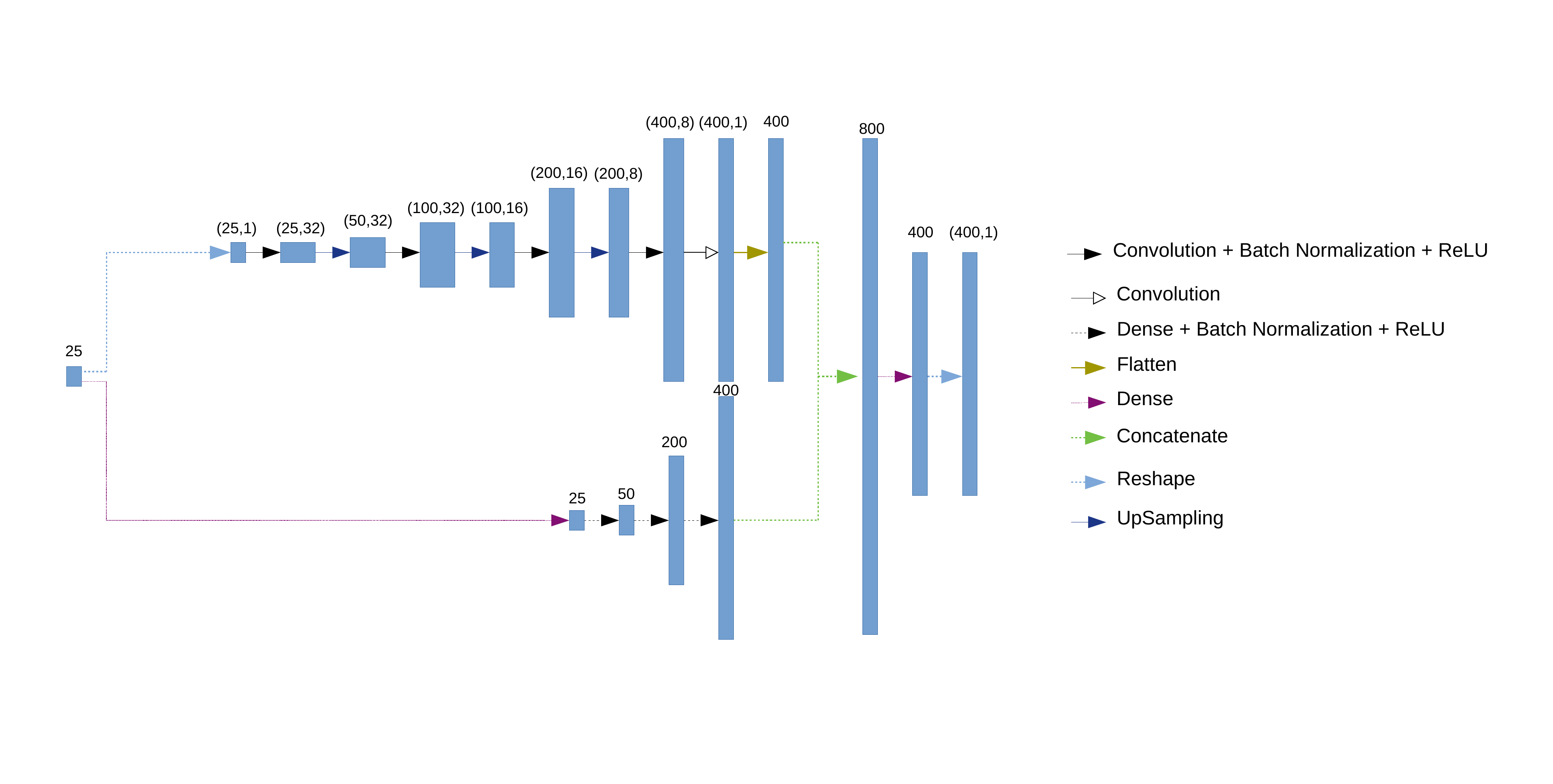}
\caption{Decoder architecture}\label{pic_decoder}
\end{figure*}

\section{Algorithm}
\subsection{Preprocessing}

Our original ECG is a $10$-second $12$-lead signal with a frequency of $500$ Hz. Each signal is cut into nine-second signals. Using the segmentation algorithms described in \cite{MoskalenkoZolotykhOsipov2019}, we determine beginnings and endings of all P and T waves and all the picks R. 
Next, we do the step forward and backward from the R pick at an equal distance.  
Thus, we obtain the set of cardiac cycles, each of which of length $400$.

\subsection{Neural network architecture. Encoder}

A variational autoencoder \cite{KingmaWelling2013,Doersch2016} consists of an encoder and a decoder.  
We propose the followin architecture for them.
The encoder  consists  of  a convolutional  and a fully connected blocks.
The architecture of the encoder is presented in Figure~\ref{pic_encoder}.
The input vector of length $400$ is fed to the input of the encoder.
Next, there is a branching into a fully connected and convolutional chains. 


The convolutional chain (at the top of the circuit in Figure~\ref{pic_encoder}) consists of $4$ series-connected blocks, each of which consists of a convolution layer, a batch normalization layer, a ReLU activation function and a MaxPooling layer. 
Next, we have another convolution layer. At the output of this block we get $25$ neurons. 

The fully connected chain of the encoder (at the bottom of the circuit in Figure~\ref{pic_encoder}) consists of $3$ fully connected (dense) layers, interconnected by a batch normalization and ReLU activation functions. At the output of the last fully connected layer we have $25$ neurons. 

The outputs of the convolutional and fully connected chains are concatenated, which gives us a vector of length $50$. Using two fully connected layers we get two $25$-dimensional vectors which interpreted as a vector of means and a vector of logarithms of variances for $25$ normal distributions (or for one $25$-dimensional normal distribution with a diagonal covariance matrix). The output of the encoder is a vector of length $25$ in which each component is sampled from those normal distributions with specified means and variance.

We will interpret this $25$-dimensional vector as a vector of new features sufficient to describe and restore with small error the one cardiac cycle. 

As the loss function, the Kullback--Leibler distance \begin{equation}\label{KL-distance}
	D_{KL}(P \parallel Q) = \int_X p\log{\frac{p}{q}} d\mu
\end{equation}
is used. Due to this fact those $25$ new features are of normal distribution.
In (\ref{KL-distance}) $\mu$ is any measure on $X$ for which there exists a function absolutely continuous with respect to $\mu$: $p = \frac{dP}{d\mu}$ and $q = \frac{dQ}{d\mu}$, $P$ is the initial distribution, $Q$ is the new distribution we have obtained.

\subsection{Neural network architecture. Decoder}

The architecture of the decoder is presented in Figure~\ref{pic_decoder}.
As an input, the decoder accepts the $25$-dimensional vector of features. Then, similarly to the encoder, branching into convolutional and fully connected chains occurs. 

The fully connected chain (at the bottom of the circuit in Figure~\ref{pic_decoder}) consists of $4$ blocks, each of which contains a fully connected (dense) layer, batch normalization layer and the ReLU activation function. 

The convolutional chain (at the top of the circuit in Figure~\ref{pic_decoder}) performs a deconvolution. It consists of $4$ blocks consisted of a convolutional layer, a batch normalization layer, and ReLU activation function, followed by and an upsampling layer. 

As a result of the convolutional and the fully connected chains, we get $400$ neurons from each. Next, we concatenate two results, obtaining $800$ neurons. Using a dense layer we get $400$ neurons which represents the ECG restored.

As a loss function for the output of the decoder, we use the mean squared error. 

\section{Experimental results}\label{AA}

\begin{figure*}
\centering
\includegraphics[width=1\textwidth]{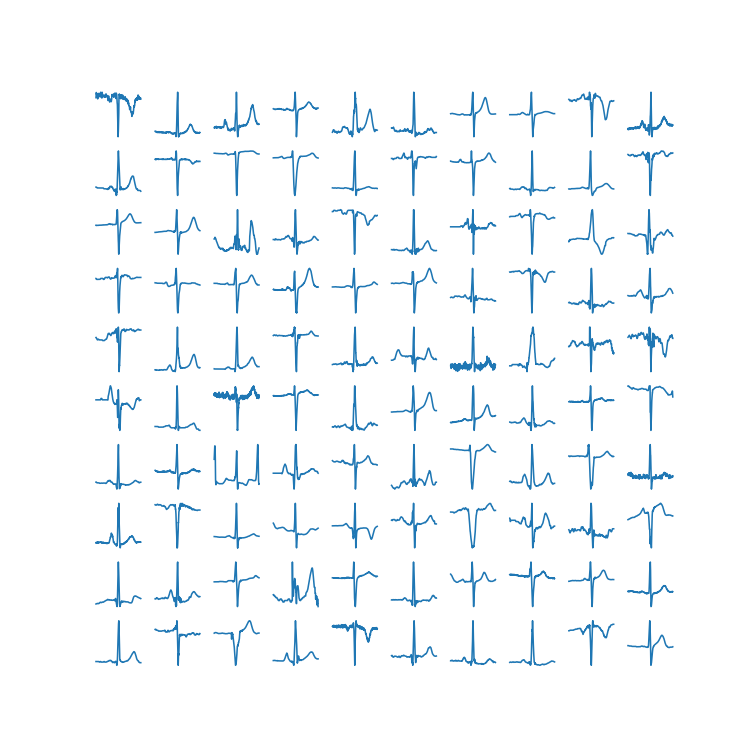}
\caption{Examples of cardiac cycles in the training sets.}\label{fig-res-real}
\end{figure*}

As a training test, we use $2033$ $10$-second ECG signals of frequency $500$ Hz \cite{Kalyakulina2018,Kalyakulina2019}. We process them according to the principles as described above and train our network on the obtained $252636$ cardiac cycles. Examples of those cardiac cycles are presented in Figure \ref{fig-res-real}.

\begin{figure*}
\centering
\includegraphics[width=1\textwidth]{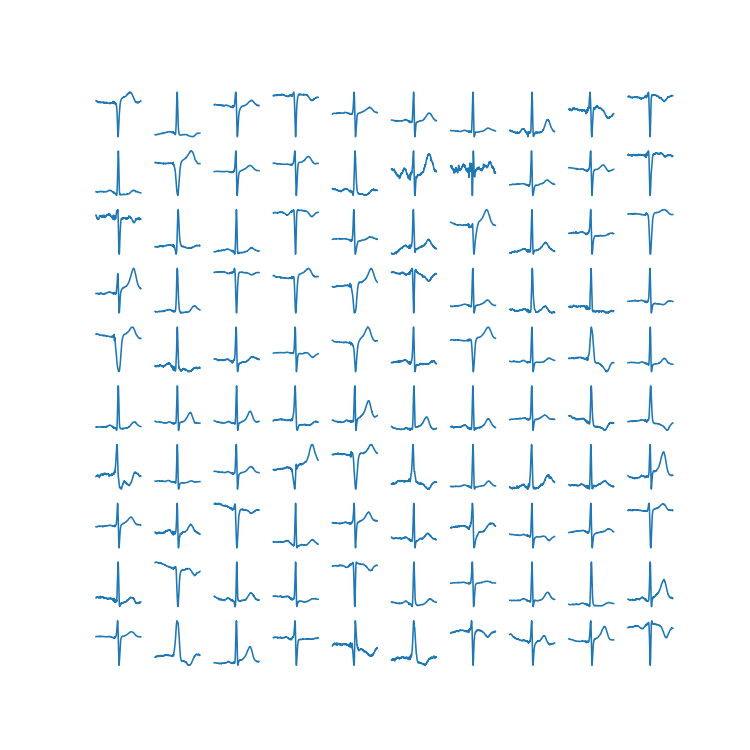}
\caption{Examples of generated cardiac cycles.}\label{fig-res}
\end{figure*}

After training the network we can test the decoder by supplying random (generated according to the standard normal distribution) numbers to its input.
The examples of the produced results are given in Figure \ref{fig-res}. These synthetic generated ECG look quite natural.

Also, for evaluating our results we calculated the Maximum Mean Discrepancy (MMD) metric (see \cite{DelaneyBrophyWard2019}) on the set of $3000$ generated ECG. The value of MMD is equal to $3.83\times 10^{-3}$. Remark that the best value of MMD obtained in \cite{DelaneyBrophyWard2019} by GAN is 
$1.05\times 10^{-3}$. However, we note that the comparison of these two metric values is not very correct, since these values were obtained on different training tests and for solving similar, but different problems.
Unfortunately, the papers \cite{ZhuYeFuLiuShen2019,GolanyRadinsky2019} don't contain (applicable to our problem) values of similar metrics.

\begin{figure}
\centering
\minipage{0.33\linewidth}
  \includegraphics[width=\linewidth]{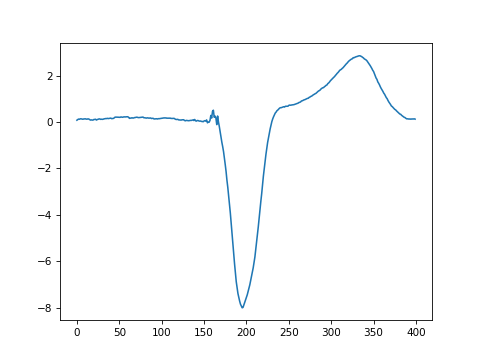}
  \includegraphics[width=\linewidth]{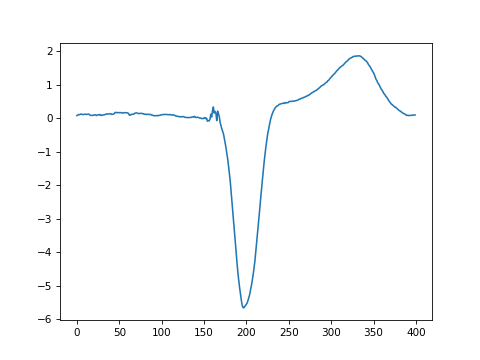}
  \includegraphics[width=\linewidth]{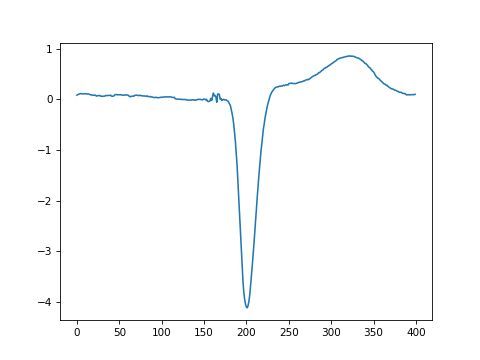}
  \includegraphics[width=\linewidth]{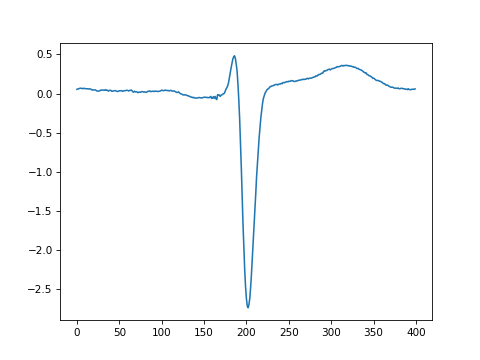}
  \includegraphics[width=\linewidth]{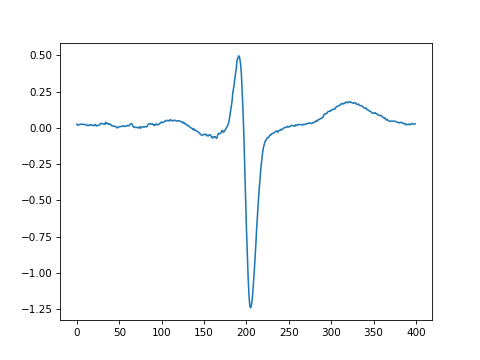}
  \includegraphics[width=\linewidth]{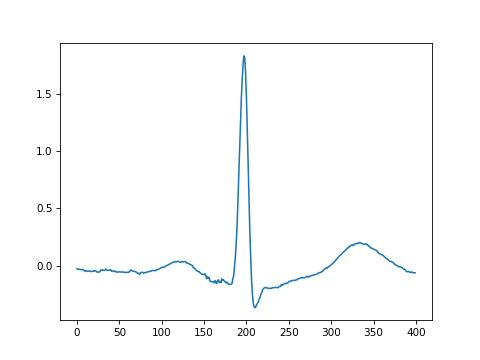}
  \includegraphics[width=\linewidth]{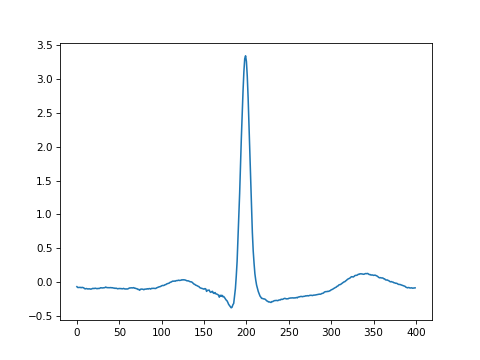}
  \includegraphics[width=\linewidth]{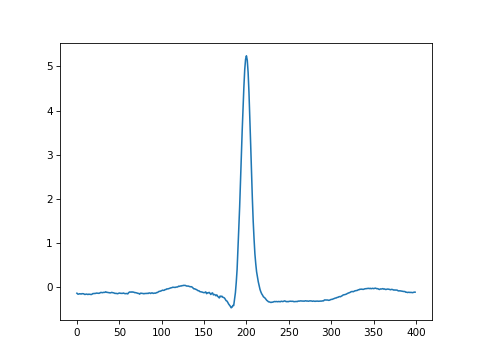}
  \includegraphics[width=\linewidth]{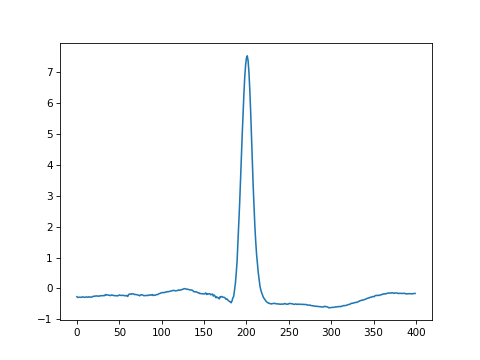}
  \includegraphics[width=\linewidth]{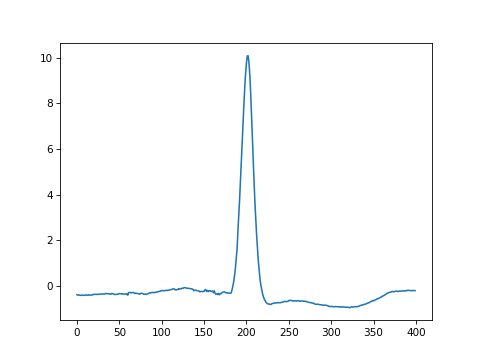}
\endminipage
\minipage{0.33\linewidth}
  \includegraphics[width=\linewidth]{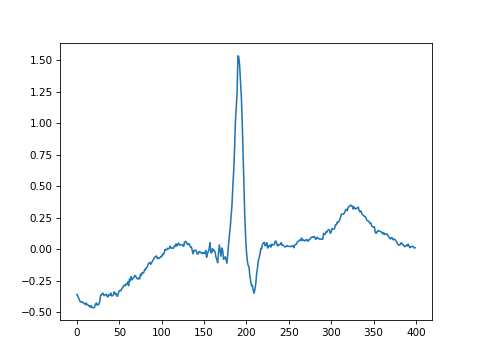}
  \includegraphics[width=\linewidth]{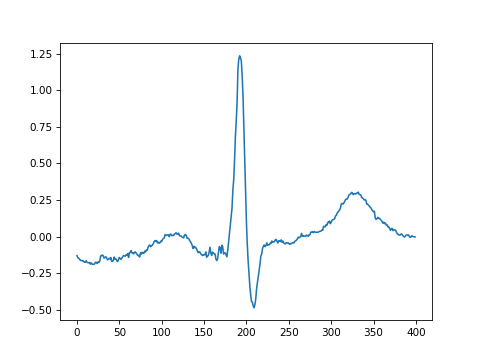}
  \includegraphics[width=\linewidth]{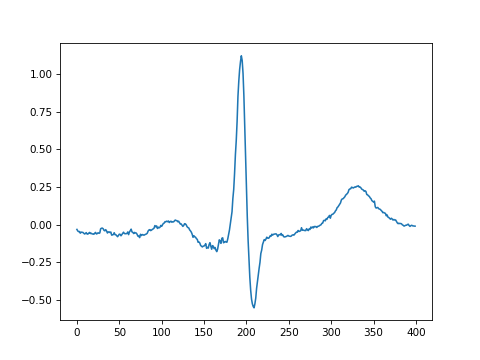}
  \includegraphics[width=\linewidth]{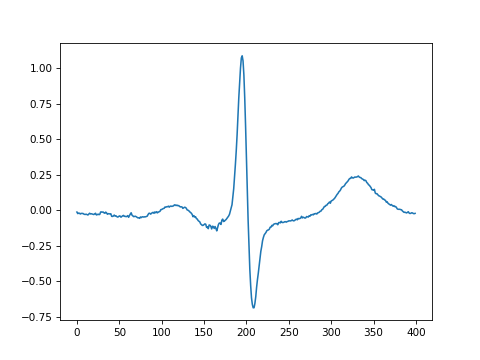}
  \includegraphics[width=\linewidth]{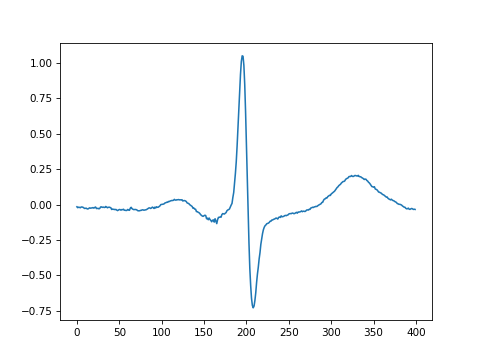}
  \includegraphics[width=\linewidth]{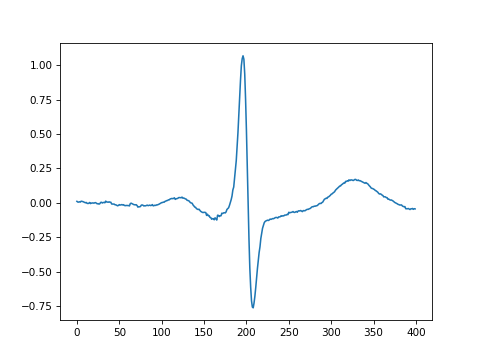}
  \includegraphics[width=\linewidth]{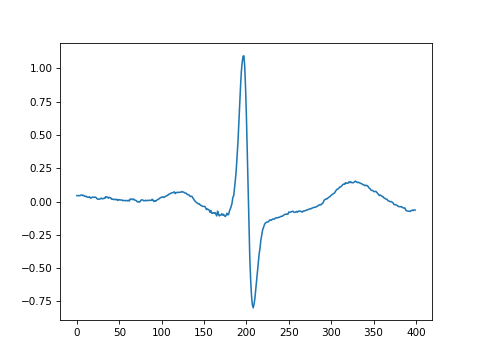}
  \includegraphics[width=\linewidth]{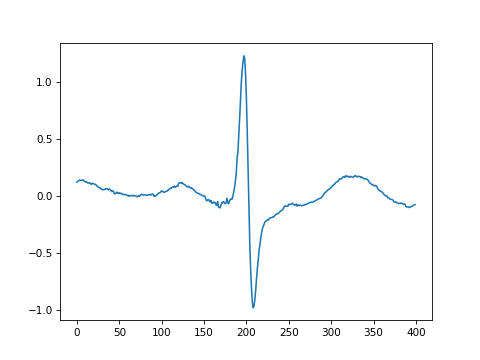}
  \includegraphics[width=\linewidth]{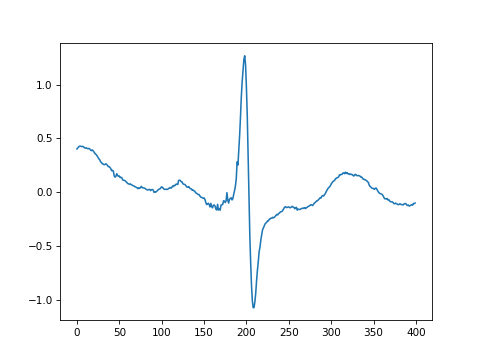}
  \includegraphics[width=\linewidth]{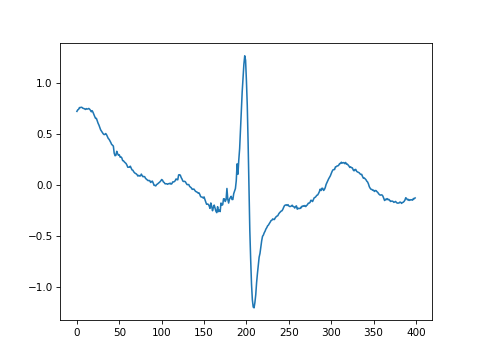}
\endminipage
\minipage{0.33\linewidth}
  \includegraphics[width=\linewidth]{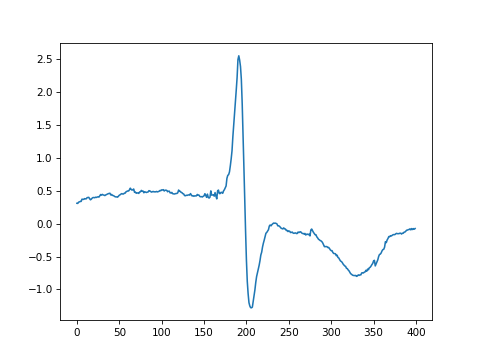}
  \includegraphics[width=\linewidth]{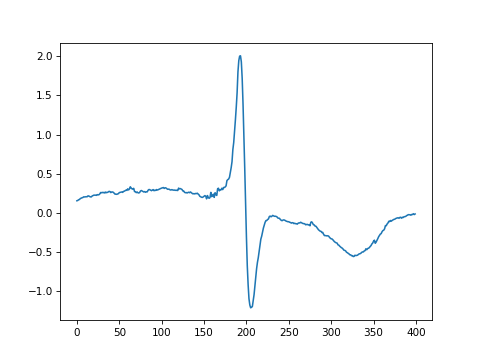}
  \includegraphics[width=\linewidth]{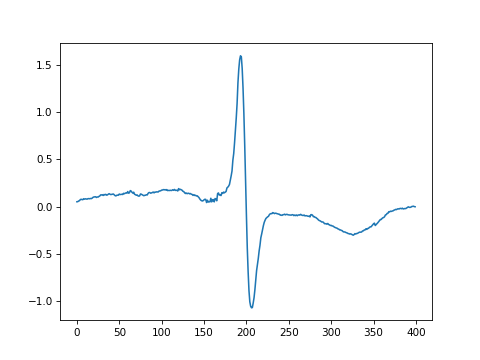}
  \includegraphics[width=\linewidth]{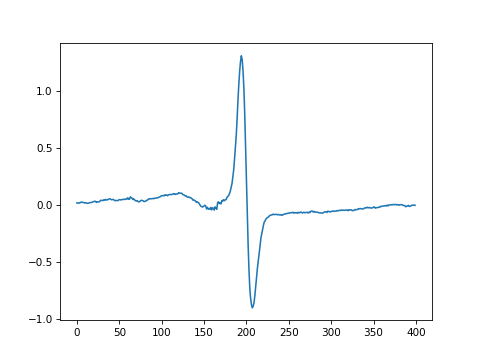}
  \includegraphics[width=\linewidth]{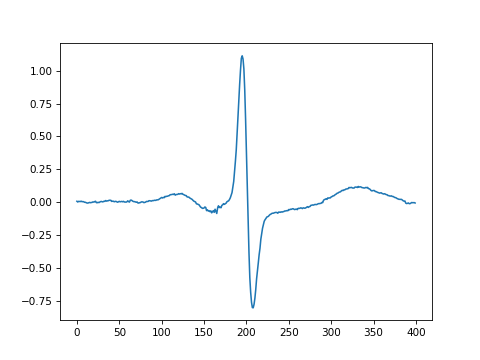}
  \includegraphics[width=\linewidth]{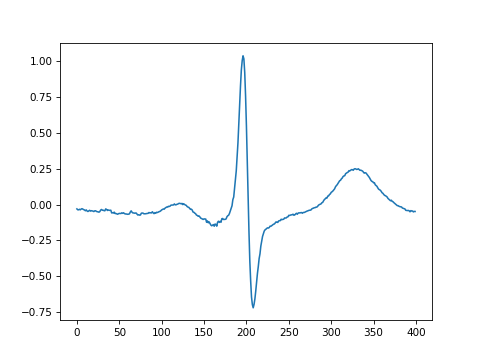}
  \includegraphics[width=\linewidth]{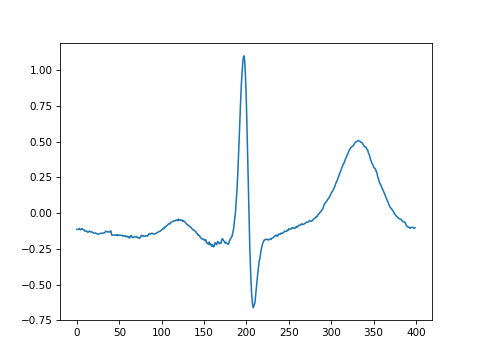}
  \includegraphics[width=\linewidth]{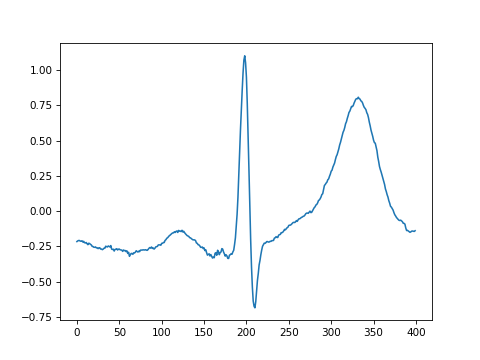}
  \includegraphics[width=\linewidth]{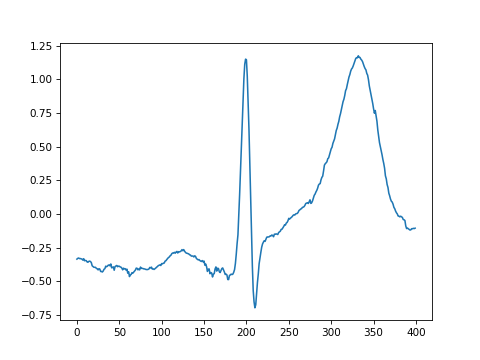}
  \includegraphics[width=\linewidth]{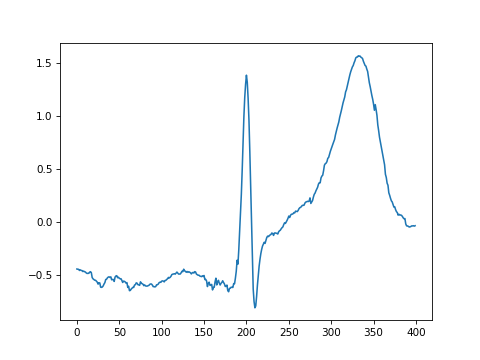}
\endminipage
\caption{Examples of ECG generated when a parameter is varying. Each column correspond to the set of fixed $24$ features and varying other feature.}
\label{fig-var-param}
\end{figure}

Interesting results were obtained when generating ECG with a varying feature. Some generated ECG signals are presented in Figure \ref{fig-var-param}. For each test, $24$ features were fixed when the remaining feature changed. It was possible to find a parameter responsible, for example, for the height of the wave T, the depression of the ST wave, etc. Thus, in many cases, the extracted features can be interpreted, which also confirms the high quality of the constructed feature description.

\section{Conclusions and further research}

In this paper, we proposed a neural network (variational autoencoder) architecture that is used to generate an ECG corresponding to a single cardiac cycle. Our method generates synthetic ECGs with completely natural appearance, which can be used to augment the training sets in supervised learning problems involving ECG. Also, our method allowed us to extract new features that accurately characterize the ECG. Experiments show that the extracted features are usually amenable to good interpretation.

We plan to use our approach to generate the entire ECG, not just one cardiac cycle. We will also use the extracted features to improve the quality of automatic diagnosis of cardiovascular diseases.


\begin{thebibliography}{99}

\bibitem{GolanyRadinsky2019}
{\it T. Golany, K. Radinsky} 
PGANs: Personalized Generative Adversarial Networks for ECG Synthesis to Improve Patient-Specific Deep ECG Classification.
Proceedings of the AAAI Conference on Artificial Intelligence, 33 (2019)

\bibitem{GoodfellowBengioCourville2016}
{\it I. Goodfellow, Y. Bengio, A. Courville} 
Deep learning. MIT press (2016)

\bibitem{Goodfellowet2014}
{\it I. Goodfellow, J. Pouget-Abadie, M. Mirza, B. Xu, D. Warde-Farley, S. Ozair, A. Courville, Y. Bengio} Generative Adversarial Networks. Proceedings of the International Conference on Neural Information Processing Systems (NIPS 2014), 2672--2680 (2014)

\bibitem{GorbanKeglWunschZinovyev2008}
{\it A. N. Gorban, B. Kégl, D. C. Wunsch, A. Y. Zinovyev (Eds.)}  Principal manifolds for data visualization and dimension reduction. Springer, Berlin (2008)


\bibitem{HongZhouShangXiaoSun2019}
{\it S. Hong, Y. Zhou, J. Shang, C. Xiao, J. Sun}. Opportunities and Challenges in Deep Learning Methods on Electrocardiogram Data: A Systematic Review.  arXiv:2001.01550 (2019)


\bibitem{DelaneyBrophyWard2019}
{\it A. M. Delaney, E. Brophy, T. E. Ward} 
Synthesis of Realistic ECG using Generative Adversarial Networks. arXiv:1909.09150 (2019)

\bibitem{Doersch2016}
{\it C. Doersch} Tutorial on variational autoencoders. arXiv preprint
arXiv:1606.05908, 2016. (2016)

\bibitem{Kalyakulina2018}
{\it A.I. Kalyakulina, I.I. Yusipov, V.A. Moskalenko, A.V. Nikolskiy, A.A. Kozlov, K.A. Kosonogov, N.Y. Zolotykh, M.V. Ivanchenko}
LU electrocardiography database: a new open-access validation tool for delineation algorithms.
arXiv:1809.03393 (2018)

\bibitem{Kalyakulina2019}
{\it A.I. Kalyakulina, I.I. Yusipov, V.A. Moskalenko, A.V. Nikolskiy, A.A. Kozlov, N.Y. Zolotykh, M.V. Ivanchenko} Finding Morphology Points of
Electrocardiographic-Signal Waves Using Wavelet Analysis. Radiophysics and
Quantum Electronics, 61(8-9), 689–703 (2019)

\bibitem{KingmaWelling2013}
{\it D. P. Kingma, M. Welling} Auto-encoding variational Bayes. arXiv preprint arXiv:1312.6114
(2019)

\bibitem{LeCunBengioHinton2015}
{\it Y. LeCun, Y. Bengio, G. E. Hinton} Deep learning. Nature,
521 (7553), 436--444 
(2015)

\bibitem{MoskalenkoZolotykhOsipov2019}
{\it V. Moskalenko, N. Zolotykh, G. Osipov} 
Deep Learning for ECG Segmentation. 
International Conference on Neuroinformatics,   Springer, Cham. 246--254 (2019)

\bibitem{SchlapferWellens2017}
{\it J. Schl\"apfer, H. J. Wellens} Computer-interpreted electrocardiograms: benefits and limitations. Journal of the American College of Cardiology, 70 (9), 1183--1192 (2017)

\bibitem{ZhuYeFuLiuShen2019}
{\it F. Zhu, F. Ye, Y. Fu, Q. Liu, B. Shen, B.} Electrocardiogram generation with a bidirectional LSTM-CNN generative adversarial network. Scientific reports, 9 (1), 1--11 (2019)




\end{thebibliography}

\end{document}